\documentclass[12pt]{amsart}

\usepackage{graphicx} 
\usepackage{natbib}
\usepackage{amsmath,amsthm}
\usepackage{fancyhdr}
\usepackage{setspace} 
\def\lsim{~\rlap{$<$}{\lower 1.0ex\hbox{$\sim$}}}

\begin{document}

\noindent {\bf The Effect of Lunar-like Satellites on the Orbital Infrared Light Curves of Earth-analog Planets}\\

%\begin{center}

\noindent Nicholas A. Moskovitz\footnote{Corresponding Author}\\
Institute for Astronomy\\
University of Hawaii at Manoa\\
2680 Woodlawn Drive\\
Honolulu, HI 96822, USA\\
Email: {\it nmosko@ifa.hawaii.edu}\\
Phone: (808) 956-6700\\
Fax: (808) 956-2901\\

\noindent Eric Gaidos\\
Department of Geology and Geophysics\\
University of Hawaii at Manoa\\
1680 East-West Road\\
Honolulu, HI 96822, USA\\

\noindent Darren M. Williams\\
School of Science\\
Penn State Erie, The Behrend College\\
Erie, PA 16563, USA\\

\noindent Running title: IR Light Curves of Planets with Large Satellites

%\end{center}

\clearpage

\begin{center}
{\textsc{Abstract}}\\
\end{center}
We investigate the influence of lunar-like satellites on the infrared orbital
light curves of Earth-analog extra-solar planets. Such light curves will be obtained by NASA's Terrestrial Planet Finder (TPF) and ESA's Darwin missions as a consequence of repeat observations to confirm the companion status of a putative planet and to determine its orbit. We use an energy balance model to calculate disk-averaged infrared (bolometric) fluxes from planet-satellite systems over a full orbital period (one year). The satellites are assumed to lack an atmosphere, have a low thermal inertia like that of the Moon and span a range of plausible radii. The planets are assumed to have thermal and orbital properties that mimic those of the Earth while their obliquities and orbital longitudes of inferior conjunction remain free parameters. Even if the gross thermal properties of the planet can be independently constrained (e.g.~via spectroscopy or visible-wavelength detection of specular glint from a surface ocean) only the largest ($\sim$ Mars-size) lunar-like satellites can be detected by light curve data from a TPF-like instrument (i.e. one that achieves a photometric signal-to-noise of 10-20 at infrared wavelengths).

Non-detection of a lunar-like satellite can obfuscate the
interpretation of a given system's infrared light curve so that it may resemble
a single planet with high obliquity, different orbital longitude of
vernal equinox relative to inferior conjunction and in some cases
drastically different thermal characteristics. If the thermal properties of the planet are not
independently established then the presence of a lunar-like satellite cannot be
inferred from infrared data, thus demonstrating that photometric light curves alone can only be used for preliminary study and that the addition of spectroscopic data will be necessary to properly characterize extra-solar Earth-like planets.

\clearpage

\section{Introduction}

Planets with a minimum mass of 5-6 Earth masses have recently been detected around low mass
stars \citep{Udry07,Rivera05} and it seems likely that
observatories such as CoRoT or Kepler will detect yet smaller planets
\citep{Gillon05}. Space-based observatories of the future will be capable of directly
detecting Earth-sized planets around other stars. Proposed missions
include a coronagraph operating at visible wavelengths (TPF-C)
\citep{Traub06}, and a large-baseline interferometer operating in the
infrared (TPF-I and Darwin) \citep{Beich06,Frid00}.  One goal of such
missions is to distinguish between planets that are Earth-like and can
support life, and those that are decidedly less so (e.g., analogs to
Mercury, Venus, or Mars).  Several techniques have been proposed to
carry out this classification.  Spectroscopy can reveal the presence
of atmospheric gases such as H$_2$O, CH$_4$ and O$_2$, which are indicative of
temperate conditions and/or biological activity
\citep{Des02}. Photometry in reflected light can reveal diurnal
(rotational) variability associated with ice, oceans, land and
vegetation across the surface of a planet if no clouds are present \citep{Ford01}.  The
specular ``glint'' from oceans might be detected as an increase
in the visible flux and polarization of reflected light at large phase angles
\citep{Williams08,McCullough08}. \citet{Selsis04} showed that orbital infrared light curves could reveal general thermal properties of terrestrial planets. \citet[][hereafter GW04]{GW04}
showed that diurnally averaged, orbital light curves at thermal
infrared wavelengths contain information about the thermal properties
of the planet's emitting layer (surface or clouds) and obliquity.  Such light curves would be
generated as a byproduct of repeated observations to confirm the
companion status and orbit of a putative planet and function as a first step towards characterization.  One finding of GW04 was that oceans or a thick atmosphere damp seasonal variations in temperature,
and that low or no variability in a planet's infrared light curve is indicative of the presence of oceans or a thick atmosphere. In conjunction with other characteristics, this is a signature of habitable surface conditions. These authors and the work presented here do not consider the effects of variable cloud cover.

The disk-averaged infrared flux of an orbiting planet can vary as it
presents different phases to a distant observer.  This phenomenon has
been observed for Jupiter-mass extra-solar planets with semi-major axes much less than 1 AU
\citep{Harrington06,Cowan07,Knutson07}. The variation in flux from a planet depends on
the diurnal pattern of outgoing infrared flux from the emitting
surface (either the top of the atmosphere, if any is present, cloud layers, or the
surface), which is controlled by the planet's thermal properties and
day length.  In general, significant day-night temperature differences
will occur only if
\begin{equation}
\label{eqn.inequality}
c\omega \lsim \left( \frac{\partial I}{\partial T} \right)_{\bar{T}},
\end{equation}
where $c$ is the heat capacity of the surface/atmosphere, $\omega$ is
the angular rotation rate, and the right hand side is the slope of the
outgoing bolometric infrared flux at the emitting surface vs.
temperature, $T$, evaluated at the mean surface temperature of the body, ${\bar T}$.  For the
Earth: $c = 8.34\times 10^7$~J~m$^{-2}$~K$^{-1}$ and $\partial I /
\partial T = 1.58$~W~m$^{-2}$~K$^{-1}$ for $\bar{T}=288$~K (however the effective emitting temperature of the Earth is 255 K). Thus equation
\ref{eqn.inequality} does not hold for the Earth: the Earth's day-night temperature
variation is small and would be for any Earth-like planet with a
rotation period much less than a year.  This is primarily due to the
high heat capacity of the ocean mixed layer which also moderates
surface temperatures over landmasses and controls the outgoing
infrared flux budget. 

The primordial rotation periods of terrestrial
planets are thought to be a stochastic outcome of the final stages of
formation by accretion of planetary embryos, and will be on the order
of hours to days \citep{Lissauer00}.  As a consequence, the
disk-averaged infrared flux from an Earth-like planet will only vary
significantly along the orbit if the planet has a non-zero obliquity
or eccentricity and hence seasons.  This was explored in GW04.

The Earth's Moon, lacking an atmosphere or oceans and having a lunar
day 29.5 times longer than the Earth, experiences a much larger diurnal surface
temperature variation.  Absence of recent geologic activity on the
Moon has allowed a regolith of impact ejecta to accumulate. This
material is optically dark [the average lunar Bond albedo is 0.07 \citep{Lane73} compared to the
Earth's 0.31] and has a relatively low heat capacity [that of the Moon
is $4\times 10^4$~J~m$^{-2}$~K$^{-1}$ at 29.5~days or 0.1\% of the
Earth, \citep{Muller98}].  As a result, the inequality of Equation
\ref{eqn.inequality} is satisfied and thus the Moon makes a significant
or even dominant contribution (depending upon viewing geometry) to the
variable component of the infrared flux from the Earth-Moon system.

The Moon is thought to have accreted from a circumterrestrial disk of
ejecta generated by the impact of a Mars-sized body
\citep{Hartmann86}. The high ambient temperatures and low gravity in
the transient disk explain the Moon's lack of volatiles
\citep{Pritchard00}. Current scenarios for the final stages of
terrestrial planet formation include such giant impacts
\citep{Canup98} and the results of numerical simulations suggest that they are not rare
\citep{Ida97}. Thus large satellites lacking atmospheres or oceans
may be common around extra-solar rocky planets.  Like the Moon, these
satellites would have originally formed closer to their parent planets
and their rotations would have quickly synchronized to their orbits
\citep{Gladman96,Canup98}. As a result their diurnal temperature
variation could be significant. We note that Moon-size satellites could retain an atmosphere against gravitational escape over Gyr time-scales if one was originally present.

Satellites around extra-solar planets will be unresolved by even the
most ambitious planet-finding mission: the angular separation of the
Earth and Moon at a distance of 10~pc is 0.25~mas. However, a large
satellite might reveal itself by a significant variation in the total
(bolometric) flux from the system.  Such an interpretation requires
independent knowledge of the thermal and rotational properties of the parent planet, which can be established using
spectroscopy \citep{Des02,Selsis04} or optical light curve data
\citep{Williams08,McCullough08}. If establishment of these gross
thermal properties leads to the expectation that infrared flux
variation would be small (Equation \ref{eqn.inequality}) then
observation of significant variation could be attributed to the
presence of a large satellite. In the absence of such auxiliary
information, however, the satellite contribution may result in an
assignment of erroneous thermal properties to the planet.

We present calculations of infrared light curves of an Earth-like
planet with a Moon-like satellite. The terms Earth-like or Earth-analog refer to specific thermal and orbital properties that represent those of the Earth (see \S\ref{sec.model}). The albedo, heat capacity and
orbital period of the satellite are set to that of the Moon (see above) while its
radius is allowed to vary. 

In \S\ref{sec.model} we describe the
details of the analytical energy balance model (EBM) used in these
calculations. We give an illustrative calculation in \S\ref{sec.curves}. We then determine the minimum radius of lunar-like satellite
that can be detected at infrared wavelengths around an Earth-analog planet (\S\ref{sec.detect}). In \S\ref{sec.error} we describe the biases in
planetary orbital properties that can be introduced by an undetected lunar-like
satellite. In \S\ref{sec.clouds} we describe the effects that low and high altitude clouds would have on our calculations and in \S\ref{sec.disc} we discuss the implications of our results.

\section{Model \label{sec.model}}

Our calculations are based on the infrared orbital light curve model
of GW04. These authors employ a linearized, analytic EBM to calculate
the infrared flux emitted by a planet.  This model assumes a single,
uniform planetary albedo and parameterizes the thermal inertia and
meridional heat transport across a planet's surface. The thermal
effect of clouds is accounted for by subtracting a correction term
from the outgoing flux \citep{Caldeira92}. The time-dependent surface
temperature distribution is described by a combination of Legendre
polynomials and a Fourier series that are solutions to a diffusion
equation with periodic temporal boundary conditions in a spherical
coordinate system. The disk- and diurnally-averaged infrared flux for
a prescribed viewing geometry is calculated along an entire orbit. As long as Eqn. \ref{eqn.inequality} is satisfied, then diurnally averaging the infrared flux justifies the use of a single, average planetary albedo.

We consider only one set of Earth-like planetary parameters.  Although such properties will undoubtedly vary amongst
extra-solar planets, those planets with thermal properties similar to
the Earth will be the most compelling targets of investigation. The
thermal inertia of the surface ($8.34 \times 10^7$~J~m$^{-2}$~K$^{-1}$)
and heat diffusion coefficient (0.38~W~m$^{-2}$~K$^{-1}$) were chosen
so that with an albedo ($A = 0.3055$), orbital semi-major
axis (a = 1~AU), and eccentricity (e = 0.0167) of Earth, the model reproduces
the meridional surface temperature distribution of the Earth as well
as the seasonal temperature variation at several latitudes (GW04).
 
The orbital properties ($a$, $e$ and $i$) of a real planet can be
determined by imaging or astrometry, but the
thermal properties of the planet may not be uniquely determined by
independent means. We discuss this scenario in \S \ref{sec.disc}.
Under the conditions of known orbital and thermal properties, the light curve is a function of the planet's
obliquity ($\delta_0$), the orbital longitude of inferior conjunction
relative to the spring equinox ($L_{0}$), and the orbital longitude of
the apastron ($L_{ap}$). If the orbit of the planet is nearly circular
then the longitude of apoastron $L_{ap}$ (fixed here to $180^{\circ}$)
is unimportant.

Ocean and atmospheric circulation and the thickness of an ocean's mixed
layer may differ for Earth-like planets with obliquities that are significantly larger
than 23.5$^{\circ}$. Thus, the actual light curves of such planets would differ from those calculated with Earth-like thermal properties. We examined this effect by
comparing general circulation model (GCM) runs for $\delta_0 = 85^\circ$ to our EBM calculations (Figure \ref{fig.gcm}). The GCM used was the three-dimensional GENESIS 2 model \citep{Williams03} and the calculations were performed with $L_0 = 120^\circ$. The amplitude and general shape of the
GCM light curves are nearly identical to those of the EBM. However, we
find that the phase of the GCM light curves tend to lead those of the EBM by
$\sim35^{\circ}$, with the greatest differences occurring when the geometry of the system is such that the poles of the planet are pointed towards the observer (e.~g. high obliquity and high inclination). This is likely due to the inclusion of polar sea ice in the GCM. 

The exact origin of the offset in phase is
uncertain, but we suspect that it involves effects of seasonal changes
in cloud cover, sea ice or ocean circulation not included in the
EBM. In \S \ref{sec.detect} we show that artificially adjusting the
phase of the EBM light curve (for $i=60^\circ$) to better match the GCM calculations only
slightly increases the probability of satellite detection. In \S
\ref{sec.error}, we find that this phase lag does not affect the
conclusion that an Earth-like planet with a lunar-like satellite produces light
curves resembling those of an isolated planet with high
obliquity (i.e. one with large amplitude, see GW04 for for a comparison of light curves from Earth-like planets with high and low obliquities). We are interested only in estimating the detectability of
satellites and their gross effect on the interpretation of infrared
light curves, rather than on detailed inferences about the climates of
planets themselves and thus use the EBM to efficiently calculate light
curves over a range of obliquity values.

Our formalism for calculating the outgoing infrared flux from a lunar-like
satellite is also based upon the analysis in GW04.  In the absence of an
atmosphere or oceans, the energy-balance equation governing the
temperature ($T$) at a given point on the surface of a satellite with
no latitudinal heat transfer is
\begin{equation}
\label{eqn.energybalance}
c\frac{\partial T}{\partial t} = S \cdot (1-A) - I(T),
\end{equation}
where time is denoted by $t$, incident stellar
flux by $S$, albedo by $A$, and outgoing infrared flux $I(T)$. $S$ and
$I(T)$ are calculated as functions of longitude and latitude on the
surface of the satellite taking into account projection effects. To
analytically solve this equation three assumptions are made. First $c$
is assumed constant in time and across the satellite surface. Second,
the temperature dependence of the outgoing infrared flux is
approximated as a linearized blackbody:
\begin{equation}
\label{eqn.infrared}
I(T) = I(\bar{T}) \cdot \left(1 + 4(T-\bar{T})/\bar{T}\right).
\end{equation}
This follows the approach of classical energy balance models
\citep{North81}.  Finally, we assume tidally-locked, synchronous
rotation as is expected for large, collisionally-formed satellites
\citep{Gladman96,Canup98}. These assumptions allow Fourier series
solutions to Equation \ref{eqn.energybalance}:
\begin{equation}
\label{eqn.solutions}
T(\theta,\ell) = T_0(\theta) + \displaystyle \sum_{n=1}^N
[a_{n}(\theta) \cos(n \ell) + b_{n}(\theta) \sin(n\ell)],
\end{equation}
where $T_0$ is the mean temperature for a given latitude $\theta$ and
$\ell$ is the longitude on the surface of the satellite. $N$ is set to
10 as numerical tests show that larger values do not significantly
change the final light curve. Substitution of Equations
\ref{eqn.infrared} and \ref{eqn.solutions} into Equation
\ref{eqn.energybalance} yields expressions for the Fourier
coefficients:
\begin{equation}
a_{n}(\theta) = \frac{a_n'(B+\alpha) -
b_n'\alpha}{\alpha^2+(B+\alpha)^2}
\end{equation}
\begin{equation}
b_{n}(\theta) = \frac{a_n'\alpha -
b_n'(B+\alpha)}{\alpha^2+(B+\alpha)^2}
\end{equation}
where $a_n'$, $b_n'$, $B$ and $\alpha$ are
\begin{equation}
a_n' = \displaystyle \int^{2\pi}_0
\frac{S(\ell,\theta)}{\pi}\cos(n\ell)\,d\ell
\end{equation}
\begin{equation}
b_n' = \displaystyle \int^{2\pi}_0
\frac{S(\ell,\theta)}{\pi}\sin(n\ell)\,d\ell
\end{equation}
\begin{equation}
B = \frac{4I(\bar{T})}{\bar{T}}
\end{equation}
\begin{equation}
\alpha = \frac{\omega c}{2}\sqrt{n}.
\end{equation}
The infrared emission from the surface is calculated using
Equation \ref{eqn.infrared} and the total signal is determined by
geometric projection of the hemisphere facing the observer.

For simplicity we assume that the orbit of the satellite is coplanar
with that of the planet's orbit around the star and, because of synchronous rotation, the
satellite has zero obliquity.  Thus any variation in outgoing flux
from the satellite is due to its finite heat capacity. The
disk-averaged flux of the satellite is then independent of the
location on its orbit around the planet and depends only on the
geometric angle described by the star, satellite, and distant observer
(Figure \ref{fig.cartoon}). Variability in the satellite signal is due to its observed phase, which changes with the orbital period of the planet. In \S \ref{sec.detect} we consider
satellites that differ in size (but not surface properties) from the
Moon.

A satellite larger than the Moon will retain heat for a longer time and be more likely to have active volcanism. This could make the satellite darker, as in the case of the lunar Mare. However, with an average albedo of 0.07 the Moon is already quite dark. Fresh basalt from active volcanism on the surface of a larger satellite would have little effect on its light curve.

\section{Example Light Curves and Observations \label{sec.curves}}

Figure \ref{fig.cartoon} illustrates how an Earth-Moon analog would
appear at five evenly-spaced points in the system's orbit.  Figure
\ref{fig.curves} plots the infrared light curves produced by this
system. The bottom panel displays the disk-averaged flux, while the
top shows light curves normalized to their respective means. This
normalization (which is employed for all subsequent analysis) removes
the radius of the planet as a degree of freedom in the model.  The
calculations were performed assuming an Earth ``twin" ($\delta =
23.45^{\circ}$) with a satellite of radius, orbital period and albedo
equal to that of the Moon (0.273 R$_\oplus$, 29.5~ days and 0.073
respectively).  For these and all simulations the coplanar orbits of
the planet and satellite are inclined by 60$^\circ$ with respect to
the plane of the sky (the median value of an isotropic distribution).
The dotted line in both panels is the contribution from the planet
alone.  The peak-to-peak amplitude of the satellite's signal (dash-dots
in the bottom panel) is 55 W m$^{-2}$, whereas the planet's flux alone
varies by only 4 W m$^{-2}$. Because the thermal inertia of the
satellite is low, it displays a larger relative infrared flux
variation than that of the planet (Equation
\ref{eqn.inequality}). However, because the satellite is smaller, the
majority of the average flux originates from the planet.

The assignment of erroneous properties to a planet with an undetected
satellite is illustrated in the top panel of Figure \ref{fig.curves},
where the dashed line is the best-fit planet-only model to the
planet+satellite observations.  This model is of a planet with Earth-like thermal properties and
$\delta_0 = 75^{\circ}$ and $L_0 = 90^{\circ}$, i.e.  quite
different than the input values.  The phase and amplitude of the light
curve produced by these models are not independent and each depends on
the obliquity and orbital longitude of inferior conjunction. Adjusting
the orbital longitude of inferior conjunction could produce better
agreement with phase, but would unacceptably decrease the amplitude of the light curve.  In general large light curve amplitudes like those
produced by Earth-like planets with Moon-like satellites can only be
mimicked by single planets with very high obliquities and orbital
geometries where the northern or southern hemisphere is pointed
towards the observer.

An example of an observation scheme of five evenly-spaced measurements
is shown in the top panel of Figure \ref{fig.curves}.  In practice a minimum of
three observations are required to confirm planetary status and reject
background sources \citep{Beich06}. The TPF mission would conduct a
minimum of 3-5 observations on each star during the first two years of
a five-year mission. The remaining time would be spent on
spectroscopic follow-up of a few dozen planets \citep{Beich06}.
Spectroscopic observations could, in principle, be split into multiple
integrations, however these would preferentially occur when the planet
was near maximum elongation from the star, thus maximizing the S/N.
Both a nulling interferometer (TPF-I, Darwin) or a coronographic
imager (TPF-C) will obscure planets along some parts of orbits in the
habitable zone \citep{Brown04}. We thus consider two observing
scenarios, one consisting of five observations at equal longitudinal
intervals around the orbit, and a second consisting of 14 observation
points restricted to half of the orbit furthest from the
star.  For example, 14 $\times$ 2-day integrations of 36 high-priority
targets might be obtained in three years.  Targets in the habitable
zone of solar-mass stars (0.9-1.3~AU, \cite{Kasting93}) will have
orbital periods of 300-550 days and thus the accessible part of an
orbit can be completely observed.  For the 5-point ``confirmation"
observation scheme we assume a S/N ratio of 10 per observation, the
median value of the S/N amongst all 234 stars for which S/N $>$ 5 is
achievable for an Earth-sized planet in a 24-hour integration time. For
the 14-point ``characterization" scheme we assume a S/N of 20, which
will be the case for Earth-sized planets around the nearest 20\% of the
target stars \citep{Beich06}.

\section{Satellite Detection Limits \label{sec.detect}}

We determine the minimum size of a Moon-like satellite that can be
detected around an Earth-like planet whose gross thermal properties are outlined in \S\ref{sec.model}. We presume that the inclination
of the planet's orbit with respect to the plane of the sky is
independently measured by astrometry and use a single value of
60$^{\circ}$.  For these calculations the orbital and thermal
parameters of the planet and the orbital period, albedo and thermal
inertia of the satellite (assumed to be equal to that of the Moon) are
held fixed, while $\delta_0$, $L_0$ and the satellite radius ($R_s$)
are allowed to vary. With the exception of cases with extremely
high planetary obliquity where most of the planetary signal is at
twice the orbital period (GW04), the signal from the planet and
satellite will have the same period.  There are then three unknowns
($\delta_0$, $L_{0}$ and $R_s$) but only two measurable quantities; the
amplitude and phase of the orbital signal. Thus it is not possible to
uniquely disentangle the planetary and satellite signals.  Instead, we
define a satellite ``detection" as the case where the observations
cannot be accounted for by a planet-only model.

Our detection analysis is as follows: We generate an array of
planetary light curves over the full ranges of $\delta_0$
[0-90$^{\circ}$] and $L_{0}$ [0-360$^{\circ}$]. The satellite light
curve for a given $R_s$ is calculated and added to each planet light
curve in the array.  Each total light curve is sampled at $N$ specified
points according to either of the observing schemes described in \S
\ref{sec.curves}. Random noise with a given RMS is added to these
measurements. We then perform an exhaustive search of {\it
planet-only} light curves to find the minimum $\chi^2$-fit to the
measurements. The analysis is repeated for different values of
$R_s$. We interpret the confidence level $C$ associated with the value
of $\chi^2$ and the number of degrees of freedom as the probability
that the deviation from the planet-only model is due to the presence
of a satellite.  This is because $1-C$ is the probability that
measurements of the light curve of the planet alone would result in a
fit with a $\chi^2$ larger than the observed value, i.e. a false
positive. We set the effective number of degrees of freedom to the
minimum $\chi^2$ value associated with a best fit to N observations of
the {\it planet-only} light curve. Although this minimum $\chi^2$
value is sensitive to the random noise characteristics of a given
sampling, we find that a more robust calculation of the true $\chi^2$
minimum (i.e. averaging over a large number of random noise patterns)
produces results that are in close agreement with this single noise
characteristic approximation. Our minimum $\chi^2$ approximation produces an increase
of nearly an order of magnitude in computational efficiency over the
robust method.

We average $C$ over $L_0$, which cannot be independently determined
for any system, but will have a uniform probability distribution. In
addition, $C$ is averaged over all possible phases of the measurement
scenarios, i.e., the longitudes at which the planet is observed are
shifted by 2 degree increments over the orbital phase range $2\pi/N$.
This produces an average probability of detection $\langle C\rangle$
for a given $R_s$ and $\delta_0$. The planet's obliquity cannot be
independently determined from the light curve of the planet+satellite,
nor can it be assumed to have an isotropic distribution
\citep{Atobe07}.

In Figures \ref{fig.prob510} and \ref{fig.prob1420} we plot $\langle
C\rangle$ vs. obliquity for several values of satellite radius (in
units of planetary radius).  For an Earth-sized planet, this range of
radii corresponds to Vesta- to Mars-sized bodies.  The two figures
correspond to the ``confirmation" (5 evenly-spaced observations at S/N
= 10) and ``characterization" (14 observations at S/N = 20 distributed
around 50\% of the planet's orbit) observation scenarios described in
\S \ref{sec.curves}. As the radius of the satellite increases the
observations become increasingly inconsistent with a planet-only light
curve, thus the probability of satellite detection increases. For
satellites smaller than 0.33 planetary radii the probability of
detection is lower in the ``characterization" observing scheme. This
is an effect of the observations being distributed around only 50\% of
the orbit. The incomplete phase sampling of this scheme does not
capture the peak of the satellite flux which occurs at the longitude
of superior conjunction. However for large satellites ($R_s \geq 0.33$
planetary radii) the amplitude of the net light curve becomes so great
that the satellite is detected even without complete phase
coverage. The scatter in these probability curves is due to the
stochastic noise added to each of the sample measurements. For both
observation schemes $<C>$ is very weakly dependent on obliquity. These
results show that a Moon-like satellite (0.27 Earth radii) would only
be detectable with $\sim$30\% confidence by either observation
scheme. For the ``confirmation" and ``characterization" observing
schemes a satellite would have to be 0.5 and 0.38 planetary radii
respectively to be detected with 90\% confidence. This corresponds to
approximately Mars-sized satellites in orbit around an Earth.

In Figure \ref{fig.prob1420.gcm} we consider how these results would
change if a full three-dimensional climate model were used to generate
the planetary light curves. As previously stated, our EBM produces
light curves that lag by $\sim35^\circ$ relative to those of the
GENESIS 2 GCM for obliquities of $23.5^\circ$ and $85^\circ$. To mimic
the results of the GCM we offset the phase of the EBM
light curves by $35^\circ$ and repeat the analysis of Figure
\ref{fig.prob1420}. It would be computationally prohibitive to generate GCM light curves for the full range of obliquities that are included in this analysis, thus we approximate the GCM by applying this offset. We find that the probability of detection actually increases
slightly with the phase-adjusted pseudo-GCM light curves (Figure
\ref{fig.prob1420.gcm}). Thus, we conclude that our EBM results are
conservative estimates for the probability of satellite detection.

\section{Errors Introduced by an Undetected Satellites  \label{sec.error}}

As we showed in Figure \ref{fig.curves}, the application of
planet-only models to a set of observations can result in
mischaracterization of the Earth-like planet if a large satellite is present.
Even if the gross thermal properties of the planet have been
independently established, this will still produce erroneous values of
$\delta_0$ and $L_{0}$.  We describe this effect by recording the
"true" ($\delta_0$, $L_0$) pair for a given planet+satellite light
curve and the ($\delta_0$, $L_0$) pair of the best fit planet-only
light curve as determined by a $\chi^2$ analysis. In some instances, a
planet+satellite light curve can be fit with equally low $\chi^2$ by
more than one planet-only light curve. For these cases, we choose the
($\delta_0$, $L_0$) pairs that are closest to the true value. Figure
\ref{fig.vectors} plots the direction and proportional magnitude of the
error introduced by satellite confusion (The length of the vectors have been reduced for clarity). These simulations were run for an Earth-like planet with a satellite of radius equal to that of
the Moon and (for clarity) no intrinsic noise added to the sample
points.  In nearly all cases, the presence of a lunar-like satellite makes an Earth-analog planet
appear to have high ($> 80^\circ$) obliquity. Shifting the planetary
light curve phases by $35^\circ$ to make them resemble the GCM results
does not alter the tendency of the best-fit solutions towards high
$\delta_0$.

The vectors in Figure \ref{fig.vectors} converge towards two values of
$L_0$ ($90^\circ$ and $270^\circ$) because the planet-only light curve
is in phase with that of the satellite at these $L_0$. For high
obliquity planets, $L_0$ values of $90^\circ$ or $270^\circ$ translate
to a planetary configuration in which the southern or northern
hemispheres (respectively) face the observer. At high $\delta_0$,
these are the only geometries that produce planet-only light curves with large amplitudes
that can reasonably fit a planet+satellite system. Adjusting
for the phase difference between the EBM and the GCM does effect these
general trends in $L_0$. Instead of converging to $L_0$ values of
$90^\circ$ and $270^\circ$, the GCM vector field reaches convergence
at $L_0$ values that are shifted by $35^\circ$.

\section{The Effect of Clouds  \label{sec.clouds}}

Clouds play an important role in the energy balance and climate of the
Earth by reflecting sunlight and scattering and trapping long-wavelength
radiation; they would presumably do so for Earth-like extra-solar
planets as well.  Although clouds represent a mean 20 W m$^{-2}$ ($8\%$)
gain in radiation for the Earth \citep{Hartmann94} we are concerned here
only with their seasonal affect on the energy budget and
disk-integrated outgoing radiation.  These seasonal effects will be
most prominent when one hemisphere is presented to the
observer.  On the Earth, low clouds (stratocumulus) produce a
decrease in net radiation, while high clouds (cirrus) produce an
increase \citep{Hartmann94}.  

We estimate the thermal effect of seasonal variability in stratocumulus clouds on the light curve of an Earth twin.  We assume that these low-altitude clouds radiate at the same temperature as a clear atmosphere
and surface, i.e. they produce no additional greenhouse effect. We use a
cloud albedo of 0.7 \citep{Hartmann94}.  \citet{Comiso01} estimated
variation in cloud cover over the open ocean around Antarctica to vary
by only $\pm 1\%$ around a mean of 91\%.  Seasonal variation of $\pm 10\%$
around a mean of 80\% was observed over the North Atlantic \citep{Massons98}.  We adopt a 20\% seasonal variation in the mean as a
reasonable bracketing value.  To produce the global mean Earth albedo
of 0.31, a dark ocean ($A = 0.07$) must be covered with 37\% clouds. Although the average cloud cover on Earth is $\sim60\%$, only half of those are low-altitude stratocumulus \citep{Minnis02}, a value in reasonable agreement with the calculated fraction of 37\%. A $\pm$ 20\% fluctuation in the mean coverage produces an albedo
variation of 0.047.  We assume that the albedo variation is uniformly distributed over each (northern/southern) hemisphere and that it varies sinusoidally in phase with the summer solstice.  (This obviously produces a
non-physical discontinuity at the equator which is unimportant for the
purposes of estimating the magnitude of the effect of clouds).  We
examined the light curves of two cases: one in which an Earth twin is
observed at a moderate inclination ($i=60^{\circ}$) and the other in which a
high-obliquity Earth is observed on an edge-on orbit ($i=90^{\circ}$) so that one
hemisphere is seen nearly pole-on (results not shown). In both cases
the annual mean of the disk-averaged flux is slightly lower in the presence of low altitude clouds, but the
amplitude and phase of the variation is essentially unchanged.  Of course, pathological deviations
from terrestrial patterns of cloudiness are possible on planets not
quite like the Earth, but we have already shown with a GCM model (that
includes parameterized cloudiness) that in at least the high obliquity
regime, our conclusions are not significantly impacted.

For an Earth-like planet, high altitude clouds will cause a greenhouse effect whose net effect is to offset any decrease in infrared emission caused by their high albedo. Variation in the fraction of high-altitude cloud cover on time scales of less than $\sim 1$ day will be averaged over during typical integrations. We argue that variation in high cloud cover over longer time scales will be insignificant compared to estimated noise characteristics for a typical observation. High altitude clouds with a temperature of 210 K would emit $\sim 40\%$ less radiation than the surface. With typical high cloud cover fractions of $\sim 30\%$ \citep{Minnis02} and again assuming $\pm$ 20\% fluctuations in coverage, we calculate that fluctuations in outgoing infrared flux due to high cloud variability will be on the order of a few percent (40\% of surface flux $\times$ 30\% coverage $\times$ 20\% variability), which would be unresolved by observations with an optimistic S/N of 20. In a more extreme case (e.g. larger variability, greater mean surface coverage or lower emitting temperatures) the variability in a light curve due to high altitude clouds will act to further confuse its interpretation.

\section{Discussion  \label{sec.disc}}

The simulations presented here show that time-series infrared photometry by a TPF- or Darwin-like observatory would reveal only the very largest lunar-like satellites
(Mars-sized) around Earth-analog planets, and then only if these
Earth-like properties, i.e., the presence of oceans and/or a
substantial atmosphere, have been established by independent means,
e.g. spectroscopy or optical photometry. This conclusion holds for a wide range of planetary obliquity (Figs. \ref{fig.prob510},\ref{fig.prob1420},\ref{fig.prob1420.gcm}), assuming that the approximations of the EBM do not grossly misrepresent the infrared light curve of a high-obliquity Earth-like planet (Fig. \ref{fig.gcm}).

When interpreting infrared light curves, the presence of an undetected lunar-like satellite can suggest erroneous values of planetary obliquity and longitude of inferior
conjunction. In the case of a planet with high thermal
inertia, inferred values of $\delta_0$ near $90^\circ$ and $L_0$
values within $35^\circ$ of $90^\circ$ or $270^\circ$ may indicate the
presence of a lunar-like satellite.  This result is based on the assumption that the satellite and
planetary orbits are coplanar, which may be not be the case for high obliquity planets. \citet{Kinoshita93} showed that the orbit of a satellite will stay in the equatorial plane of its host planet if the secular rate of change of the planet's obliquity is slower than the precessional speed of the satellite orbital plane. Thus satellites around planets that experienced rapid changes in obliquity [possibly by collisions as in the case of Uranus \citep{Parisi97}] would stay in their coplanar orbits.

If a satellite's orbit is non-coplanar then its rotation axis will be tilted with respect to the plane of the planet's orbit. This effectively causes a non-zero obliquity for the satellite which will modify the amplitude of the satellite's light curve but will not change its period. If this non-coplanar orbit precesses then the signal will change over a time-scale of many orbital periods.

If the thermal properties of a planet are not independently
established via spectroscopy \citep{Des02}, visible-wavelength detection of glint from an ocean or significant polarization of visible reflectance \citep{Williams08,McCullough08}, then the flux from an unresolved lunar-like satellite can induce
serious errors. If the measurements are modeled with the several free
parameters (e.g.~$\delta_0$, $L_0$, $A$, $c$ and efficiency of meridional heat transport)
 then a set of planet+satellite measurements can be satisfactorily fit by a planet-only light curve. For instance, the peak-to-peak light curve
amplitude from a system with a large, unresolved lunar-like satellite around an Earth-analog
planet can be fit by a planet
with low thermal inertia and drastically different $\delta_0$ and $L_0$, implying a planet more akin to Mars than Earth. Such an erroneous
inference would impact the determination of
the frequency of habitable planets.  This reinforces the need for multiple wavelength observations including spectroscopy and photometry to disambiguate the characterization of extra-solar terrestrial planets \citep{Traub06,Beich06}. 

Where, optimistically, the thermal properties of a planet are known and its satellite is Mars-sized, the existence of a satellite
may be inferred from infrared
data. Such a discovery would provide information
about the collisional and kinematic evolution of the parent planet. In
addition, a large satellite could be a potential indicator of
habitability, as the presence of the Moon is known to stabilize the
obliquity and climate of the Earth \citep{Laskar93}.  \\

{\bf{Acknowledgments}}

This work has been supported by NASA Terrestrial Planet Finder Foundation science award NNG04GL48G. We thank the anonymous reviewers for their helpful comments.

%%%%%%%%%%%%%%%%%%%%%%%%%%%%%%%%%%%%%%%%%%%%%%%
% Bibliography 									                                                         %
%%%%%%%%%%%%%%%%%%%%%%%%%%%%%%%%%%%%%%%%%%%%%%%

%%%%%%%%%%%%%%%%%%%%%%%%%%%%%%%%%%%%%%%%%%%%%%%
% Figures																	 %
%%%%%%%%%%%%%%%%%%%%%%%%%%%%%%%%%%%%%%%%%%%%%%%

% NOTE: Figure 1 is a meant to be two columns (16 cm) wide. Figures 2-5 are single column.
%
% Figures are to be inserted into the text nearest to their point of first reference.

\begin{center}

\begin{figure}[b]
\includegraphics*[]{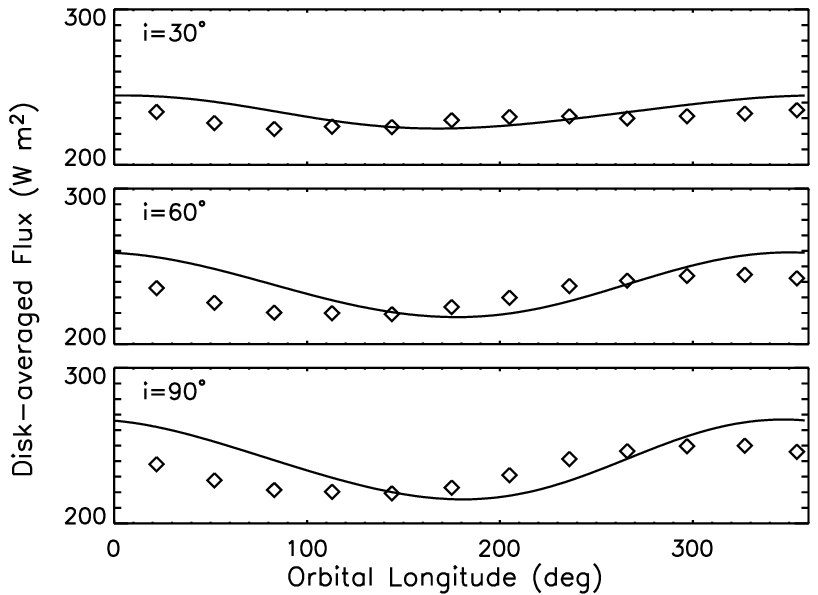}
\caption{\doublespacing Calculated EBM (solid) and GCM (diamonds) light curves of a high-obliquity ($\delta_0 = 85^\circ$) planet with Earth-like thermal properties. Light curves are shown for three values of inclination. $L_0 = 120^\circ$ (defined with respect to the vernal equinox) is used for both models. The two models are in close agreement regarding the amplitude of the planetary signal, however the phase of the EBM calculations tend to lag by $\sim 35^\circ$.
}
\label{fig.gcm}
\end{figure}

\begin{figure}[b]
\includegraphics*[width=16cm]{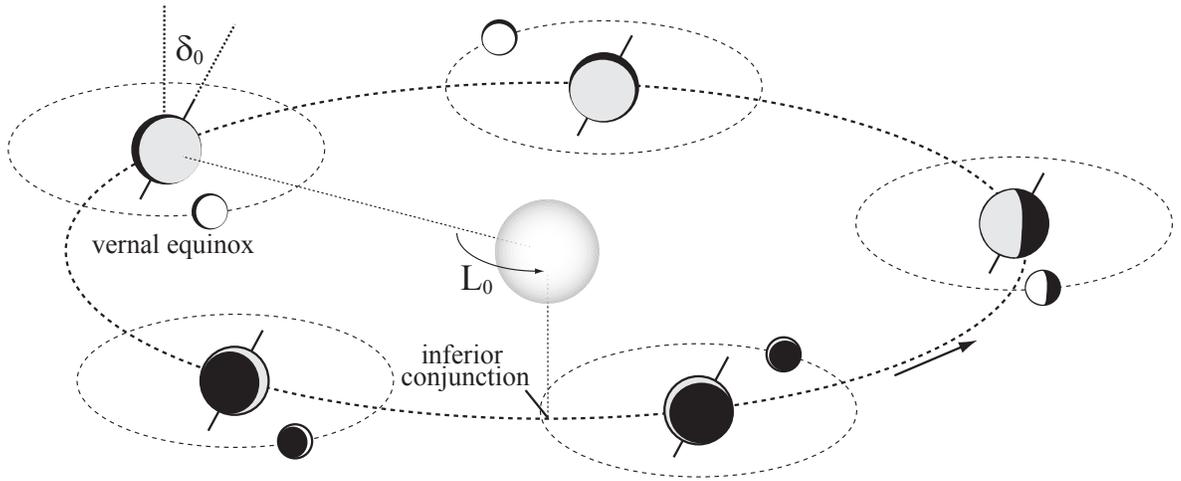}
\caption{\doublespacing Planet and satellite at
five uniformly-spaced orbital longitudes. The vernal equinox corresponds to $L=0$ and inferior conjunction to
$L=L_0$.  The disk-averaged flux from the Earth-like planet, which has
a high thermal inertia, varies only with seasonal surface temperature
differences between the two hemispheres.  The disk-average flux from
the Moon-like satellite, which has a low thermal inertia, depends on
its observed phase. The observed phase of the satellite depends only on its orbit around the star, not around the planet.
}
\label{fig.cartoon}
\end{figure}

\begin{figure}[b]
\includegraphics*[]{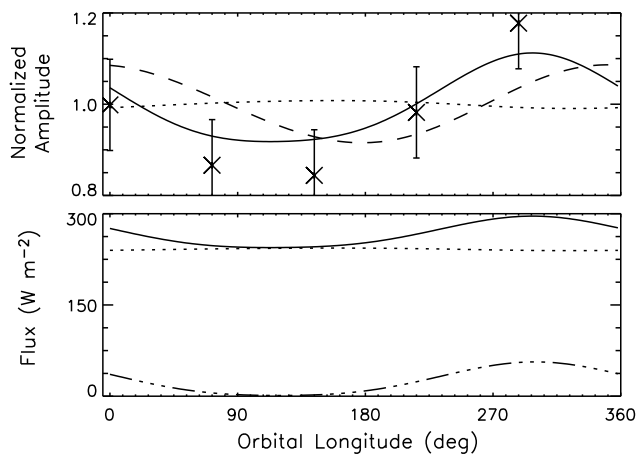}
\caption{\doublespacing Simulated light curves and observations of an Earth-Moon ``twin". The bottom panel shows disk-averaged flux while the top displays the normalized light curves. In both panels the dotted curve is the planetary seasonal flux and the solid curve is the ``true" planet+satellite light curve. The dash-dot curve in the bottom panel is the satellite contribution. In the top panel the sample points are indicated by X's with one-sigma error bars for a S/N = 10. The best fit, planet-only light curve to the sample points is represented by the dashed line. The top panel light curves have been normalized by their mean to remove any dependence on the radius of the planet.}
\label{fig.curves}
\end{figure}

\begin{figure}[b]
\includegraphics*[]{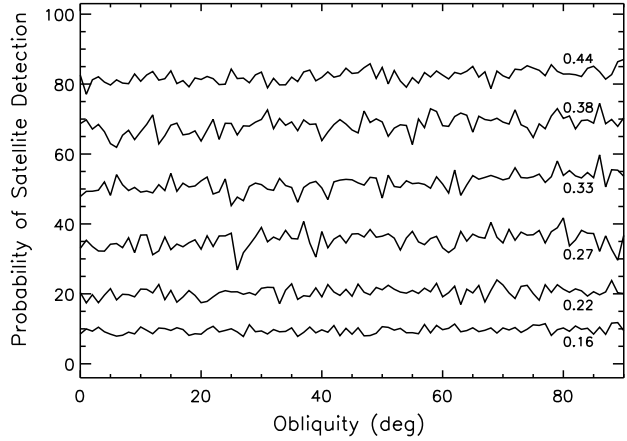}
\caption{\doublespacing The $L_0$ averaged probability of satellite detection as a function of planetary obliquity for a range of satellite radii (0.16, 0.22, 0.27, 0.33, 0.38 and 0.44 planetary radii). The radius of the Moon is 0.27 $R_\oplus$. These simulations were performed with 5 evenly-spaced observations and a S/N of 10.  As the radius of the satellite increases it becomes increasingly difficult to explain the sample measurements with a planet-only light curve, thus the probability of detection increases.}
\label{fig.prob510}
\end{figure}

\begin{figure}[b]
\includegraphics*[]{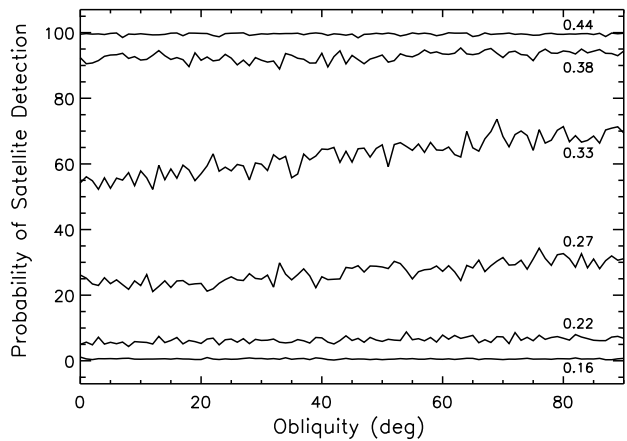}
\caption{\doublespacing Similar to Figure \ref{fig.prob510}, except for the ``characterization" observing scheme: 14 sample points with a S/N of 20, distributed around 50\% of the planet's orbit. Although the S/N is higher than the observations of Figure \ref{fig.prob510}, the distribution of sample points around only 50\% of the orbit makes it difficult to detect satellites of small radii ($\leq 0.33$ planetary radii).}
\label{fig.prob1420}
\end{figure}

\begin{figure}[b]
\includegraphics*[]{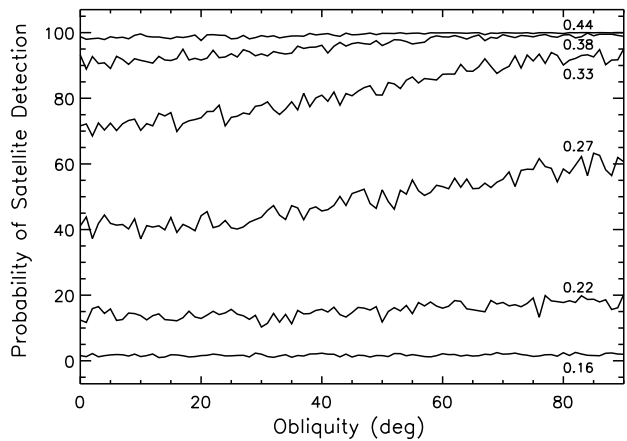}
\caption{\doublespacing Similar to Figure \ref{fig.prob1420}, except that a $35^\circ$ phase shift has been applied to all EBM planetary light curves so that they agree with the GENESIS 2 GCM. Using these pseudo-GCM light curves the detection probability actually increases relative to the EBM case (Figure \ref{fig.prob1420}). This suggests that the EBM places a lower limit on the probability of satellite detection.
}
\label{fig.prob1420.gcm}
\end{figure}

\begin{figure}[b]
\includegraphics*[]{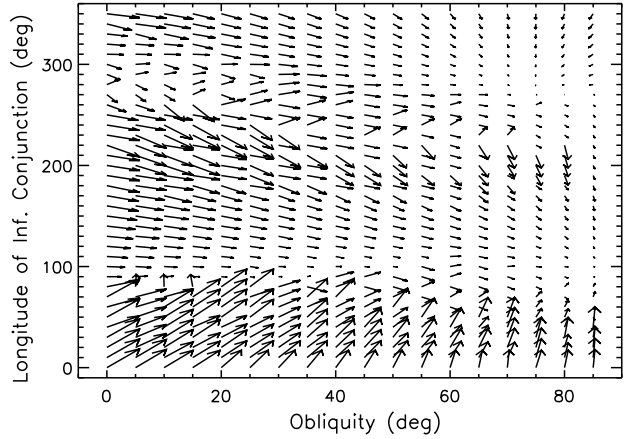}
\caption{\doublespacing Error induced in the light curve analysis by an undetected satellite. The vectors point from the true ($\delta_0$, $L_0$) pair that is used to generate the planet+satellite light curve, towards the best fit planet-only ($\delta_0$, $L_0$) pair. For clarity, the length of the vectors have been scaled to the amount of induced error. Their endpoints do not actually fall on the best-fit ($\delta_0$, $L_0$) values. Nearly all best fit values cluster at high obliquity and $L_0$ values of $\sim 90^\circ$ or $\sim 270^\circ$.}
\label{fig.vectors}
\end{figure}

%%%%%%%%%%%%%%%%%%%%%%%%%%%%%%%%%%%%%%%%%%%%%%%
% Tables											                                                         %
%%%%%%%%%%%%%%%%%%%%%%%%%%%%%%%%%%%%%%%%%%%%%%%
%\clearpage

% two column table
%\begin{table}[]
%\caption{Default Planet Properties}
%\doublespacing
%\begin{tabular}{ccccccc}
%\hline
%\hline
%{\it a} & & & & & Heat Capacity & Efficiency of Heat Transport \\
%(AU) & {\it e} & {\it i} & $L_{ap}$ & Albedo & (J m$^{-2}$ K$^{-1}$) & (W m$^{-2}$ K$^{-1}$) \\
%\hline
%1.0 & 0.0167 & $60^\circ$ & $180^\circ$ & 0.3055 & $8.34 \times 10^7$ & 0.38 \\
%\hline
%\end{tabular}
%\label{tab1}
%\end{table}

%\clearpage

% single column table
%\begin{table}[]
%\caption{Properties of Figure 2 Light Curves}
%\doublespacing
%\begin{tabular}{cccc}
%\hline
%\hline
%Top Panel Curve & $\delta_0$ & $L_{0}$ & Moon? \\
%\hline
%\doublespacing

%Solid, Planet and satellite & 23.45$^\circ$ & 120$^\circ$ & Yes\\
%Dotted, Planet only & 23.45$^\circ$ & 120$^\circ$ & {No}\\
%Dashed, Best fit planet & 75$^\circ$ & 90$^\circ$ &{No}\\
%\hline
%\end{tabular}
%\label{tab2}
%\end{table}

\end{center}

\end{document}